\documentclass[prb,aps,twocolumn]{revtex4}
\usepackage{amsmath}
\usepackage{amsfonts}
\usepackage{amssymb}
\usepackage{graphicx}

\input{tcilatex}

\begin{document}

\title{Theory of superconducting pairing near the mobility edge}
\author{M. V. Feigel'man$^1$, L. B. Ioffe$^{2,1}$ and E. A. Yuzbashyan$^2$}
\affiliation{$^1$L. D. Landau Institute for Theoretical Physics, \\
Kosygin str.2, Moscow 119334, Russia}
\affiliation{$^2$ Serin Physics Laboratory, Department of Physics and Astronomy,\\
Rutgers University, Piscataway, NJ 08854, USA}
\date{\today}

\begin{abstract}
We develop a theory of a pseudogap state appearing near the
superconductor-insulator transition in strongly disordered metals with
attractive interaction. We show that such an interaction combined with the
fractal nature of the single particle wave functions near the mobility edge
leads to an anomalously large single particle gap in the superconducting
state near SI transition that persists and even increases in the insulating
state long after the superconductivity is destroyed. We give analytic
expressions for the value of the pseudogap in terms of the inverse
participation ratio of the corresponding localization problem.
\end{abstract}

\maketitle

\renewcommand{\L}{L_{\rm loc}}

\begin{figure}[th]
\includegraphics[width=2.5in]{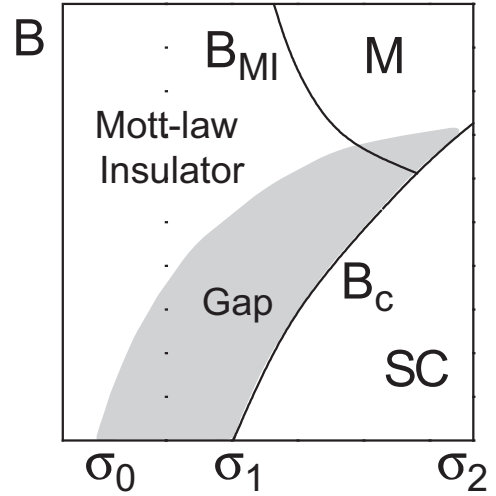}
\caption{Schematics of the phase diagram of disordered superconductors in
the vicinity of the superconductor-insulator transition as a function of
the magnetic field, $B$, and disorder strength that is characterized by  room temperature
conductivity, $\protect\sigma \sim k_{F}l$. In the shaded region
 the resistiviy exhibits activated behavior with a well defined gap $%
T_{I}$.}
\label{PhaseDiagram}
\end{figure}

A rapidly growing number of experiments~\cite%
{Shahar1992,Goldman1993,Kowal1994,Gantmakher2000,Adams2001,Baturina2004,Shahar2004,Shahar2005,Steiner2005}
on various disordered superconductors show that a novel phase appears in the
vicinity of the superconductor-insulator transition as sketched in the low
temperature phase diagram in Fig.~\ref{PhaseDiagram}. On the superconducting
side of the transition a relatively small magnetic field suppresses the
superconductivity leading to an insulator characterized by a large thermally
assisted resistance with a hard gap. Upon a further increase of the magnetic
field the resistance and the gap drop~\cite%
{Gantmakher2000,Adams2001,Baturina2004,Shahar2004,Shahar2005}. It is
tempting to explain these data by a formation of localized Cooper pairs\cite%
{Gantmakher1998} -- in this picture the superconductivity is due to a
fragile coherence between localized Cooper pairs, while the energy of single
electron excitations is much larger and remains finite even when the
coherence (and thus the superconductivity) is destroyed. Experimentally,
this behavior is observed only for a limited range of disorder strengths, $%
\sigma _{1}<\sigma <\sigma _{2}$, characterized by the conductivity $\sigma $
at room temperature. At weaker disorder, $\sigma \geq \sigma _{2}$, the
destruction of superconductivity by a magnetic field leads, as usual, to a
formation of a normal metallic state without any noticeable localization
effects. At larger disorder $\sigma <\sigma _{1}$ superconductivity is
replaced by a gapped insulator even at $B=0$. Usual insulator with a
variable-range hopping conductance is recovered at larger disorder, $\sigma
<\sigma _{0}$.

The hypothesis of preformed Cooper pairs is further confirmed by the
behavior of disordered superconductors at higher temperatures. On the
insulating side of the transition in thick (effectively 3D) films one
observes~\cite{Shahar1992,Kowal1994} Arrenius behavior, $R(T)\propto \exp
(T_{I}/T)$, at low temperatures. The experimental value of the activation
energy, $T_{I}$, is somewhat larger than the superconducting gap in less
disordered samples and grows with the disorder. For instance in InO$_{x}$ films $%
T_{I}=2-15\,K$, while the critical temperature $T_{c}=3.3\,K$ in best samples~\cite%
{Shahar1992,Kowal1994}. However, at higher temperatures ($T\geq 10K$ for InO$%
{x}$ films studied in Ref.~\onlinecite{Kowal1994}) this behavior is replaced
by Mott's variable range hopping $R(T)\sim \exp (T_{M}/T)^{1/4}$. Thus,
qualitatively the temperature plays the same role as the magnetic field,
suppressing an unconventional insulator formed near superconductor-insulator
transition. This can be easily understood if the insulating pseudogap is due
to preformed Cooper pairs with a pairing energy somewhat larger than the
superconductive gap in a less disordered samples.

The purpose of this Letter is to provide a justification for this picture
and to construct the theory of localized Cooper pairs starting from a
semi-microscopic model that contains only low energy electrons with BCS-type
attraction and a strong random potential which leads to Anderson
localization of single-particle states. An important ingredient of the
problem is the fractal structure of single-particle wave functions, which
sets in on a microscopic scale $L_{\mathrm{0}}$ determined by the disorder
potential and extends to the localization length $L_{\mathrm{loc}}$.

It is well established that disorder has little effect on conventional
superconductivity\cite{Anderson1959,Abrikosov1959} in the absence of Coulomb
repulsion. On the mean-field level (neglecting thermal fluctuations), $T_{c}$
and the self-consistent BCS gap $\Delta $ are affected by the localization and start to decrease only
when the effective level spacing in the localization volume, $\delta
_{L}=1/(\nu _{0}L_{\mathrm{loc}}^{3})$, becomes comparable~\cite%
{Anderson1959,Sad1984,Kot85,Ma1985,Larkin1999} to  $\Delta $ (here $\nu _{0}$
is the density of orbital states at the Fermi level). Intuitively, it is
clear that the thermodynamics of a strongly disordered superconductor built
on localized electron wave functions is practically the same as that of a
system separated into compartments (grains) of size $L_{\mathrm{loc}}$. In
each grain $\delta _{L}$ serves as a low energy cut-off; when it exceeds $%
\Delta $ the superconductivity is destroyed. However, the attraction between
electrons persists until $\delta _{L}$ exceeds a high energy cut-off which
is typically the Debye frequency, $\omega _{D}$. The attraction results in
pairing of electrons with opposite spins occupying the same localized state.
Assuming that the pairing interaction is local and is characterized by a
dimensionless coupling $\bar\lambda $, the pairing energy, $2\Delta _{1}$, is
proportional to the inverse volume occupied by the state, $\Delta _{1}\sim
\bar\lambda /(\nu _{0}V_{\mathrm{eff}})$. It is crucial for the following that
this volume is parametrically smaller than the 'localization volume' $L_{%
\mathrm{loc}}^{3}$ due to the fractal nature of the wave functions near
the mobility edge $V_{\mathrm{eff}}\sim L_{\mathrm{0}}^{3}(L_{\mathrm{loc}}/L_{%
\mathrm{0}})^{D_{2}}$, where $D_{2}<3$, cf.~%
\onlinecite{Parshin1999,Mirlin2002}. Then the energy scale $\Delta _{1}$ is
parametrically larger than $\delta _{L}$. Thus, the strength of the disorder
is characterized by two relevant energy scales, $\delta _{L}$ and $\Delta
_{1}$.

Here we focus on two limiting cases
\begin{eqnarray}
\Delta &\ll &\delta _{L}\ll \omega _{D}  \label{cond1} \\
\delta _{L} &\ll &\Delta \ll \Delta _{1}  \label{cond2}
\end{eqnarray}
As shown below in the first regime the single-particle gap remains large
leading to a gapped insulator. In the second regime the superconductivity
persists, but the single-particle gap $\Delta_1$ is much larger than  in
conventional superconductors. In particular, the ratio of this gap to $T_{c}$
can be anomalously large. Moreover, in the latter regime the superconducting
phase transition is pre-empted (upon $T$ decrease) by a formation of a
"pseudogaped" insulating state similar to the one observed in underdoped
cuprates~\cite{cuprates}.

We assume that a BCS-type attraction between electrons at high energy scales
$\sim \omega _{D}$ is only weakly affected by localization of electron wave
functions provided that $\delta _{L}\ll \omega _{D}$. The theoretical reason
for this is the scaling argument that says that physics at short scales and
large energies is not affected by localization as long as $k_{F}L_{\mathrm{%
loc}}\gg 1$.   This is further
collaborated by an experimental observation\cite{Shahar1992,Kowal1994} that $%
T_{c}$ changes little until one gets into a deeply localized regime. In a
fermion system with weak attraction one can leave only the pair interaction
terms in the Hamiltonian leading to the usual BCS model in the basis of
localized electron states~\cite{Ma1985}:
\begin{equation}
H=\sum_{j\sigma }\epsilon _{j}c_{j\sigma }^{\dagger }c_{j\sigma }-\frac{%
1 }{\nu _{0}}\sum_{j,k}\lambda_{jk}M_{jk}c_{j\uparrow }^{\dagger }c_{j\downarrow
}^{\dagger }c_{k\uparrow }c_{k\downarrow }  \label{Ham}
\end{equation}%
where $\epsilon _{j}$ is the single-particle energy of the state $j$ and the
matrix elements $M_{jk}=\int d\mathbf{r}\psi _{j}^{2}(\mathbf{r)}\psi
_{k}^{2}(\mathbf{r})$. In  Hamiltonian (\ref{Ham}) we will also  distinguish the off-diagonal dimensionless coupling
constant $\lambda_{jk}=\lambda$ for $j\ne k$ from the diagonal one $\lambda_{jj}=\bar\lambda$ for the reasons detailed below.

To study the model (\ref{Ham}) we need the statistics of single electron
states near the 3D Anderson mobility edge. We expect that on average
off-diagonal matrix elements, $\bar{M}_{jk}$ with $j\neq k$ have the usual
scaling $1/L_{\mathrm{loc}}^{3}$ for the states localized at distances
smaller than $L_{\mathrm{loc}}$ from each other. In other words we assume
that correlations $M_{jk}$ between wave functions of different states are
much smaller than individual "inverse participation ratios" (IPRs), $%
M_{jj}=\int d\mathbf{r}\psi _{j}^{4}( \mathbf{r})$, which have the meaning
of the inverse volumes of localized states, $V_{\mathrm{eff}}$. In fact,
extensive numerical studies \cite{Parshin1999,Mirlin2002} show that a
typical IPR, $\bar{M}$, scales as $\bar{M}\approx L_{0}^{-3}(L_{\mathrm{loc%
}}/L_{0})^{-D_{2}}$, with $D_{2}=1.30\pm 0.05$.

As discussed above, the thermodynamics of this model should be similar to a
collection of grains of size $L_{\mathrm{loc}}$ with Hamiltonians
\begin{equation}
H=\sum_{j\sigma }\epsilon _{j}c_{j\sigma }^{\dagger }c_{j\sigma
}-\sum_{j,k}g_{jk}c_{j\uparrow }^{\dagger }c_{j\downarrow }^{\dagger
}c_{k\uparrow }c_{k\downarrow }  \label{grain}
\end{equation}%
where the average level spacing is $\langle \epsilon _{j+1}-\epsilon
_{j}\rangle =\delta _{L}$, the off-diagonal part of the interaction has the
usual form for a grain, $g_{jk}=g=\lambda \delta _{L}$ for $j\neq k$, while
the diagonal part has an extra term $G$ so that $g_{jj}=g+G=\bar{\lambda}(L_{%
\mathrm{loc}}/L_{\mathrm{0}})^{-D_{2}}/(\nu _{0}L_{\mathrm{0}}^{3})$ due to
the fractal nature of the wavefunctions.  We have checked that the results described in this
Letter are the same (up to numerical factors of order one) for models (\ref%
{Ham}) and (\ref{grain}).

We begin with the insulating region (\ref{cond1}) where Cooper interaction
can be treated perturbatively\cite{Matveev1997}. In the first order
perturbation theory the minimum energy (counted from the chemical
potential), $\Delta _{1},$ required to add a single electron to the system
is
\begin{equation}
\Delta _{1}=\frac{\bar{\lambda}}{2\nu _{0}L_{\mathrm{0}}^{3}(L_{\mathrm{loc}%
}/L_{\mathrm{0}})^{D_{2}}}  \label{DeltaP3}
\end{equation}%
The coupling constant $\bar{\lambda}$ appearing in this equation is not
renormalized by Cooper loops: indeed, the Hamiltonians (\ref{Ham}) and (\ref%
{grain}) do not mix unoccupied and doubly occupied (\textit{unblocked})
states with singly occupied (\textit{blocked}) ones. Blocked states do not
participate in pair scattering and contribute only their single-particle
energy to the Hamiltonian. The additional contribution coming from the
diagonal term in (\ref{grain}) is $GN_{pairs}$, where $N_{pairs}$ is the
total number of pairs. Using the fact that the total number of electrons is $%
2N_{pairs}+N_{B}$, where $N_{B}$ is the number of blocked states and
shifting single-particle levels $\epsilon _{j}\rightarrow \epsilon _{j}+G/2$%
, we can rewrite Hamiltonian (\ref{grain}) as
\begin{equation}
H=\sum_{j\sigma }\epsilon _{j}c_{j\sigma }^{\dagger }c_{j\sigma
}-g\sum_{j,k}c_{j\uparrow }^{\dagger }c_{j\downarrow }^{+}c_{k\uparrow
}c_{k\downarrow }+\sum_{B}\left( \epsilon _{k}+\frac{G}{2}\right)
\label{grain1}
\end{equation}%
Here the first two terms act only on unblocked states, while the summation
in the last term is over decoupled blocked states. The renormalization
affects only the interacting part which is of a standard BCS form. This
statement is valid for the reduced Hamiltonians~(\ref{Ham},\ref{grain})
which contain pair-wise eigenstates only. In general, one would say that our
phenomenological Hamiltonians ~(\ref{Ham},\ref{grain}) contain \textit{two
different} coupling constants originating from the competition between
Coulomb and electron-phonon interactions -- standard BCS coupling $\lambda $
and the additional "diagonal" one, $\bar{\lambda}$, which characterizes
interaction of two electrons in the \textit{same} localized state.

Applying the same arguments to the Hamiltonian (\ref{Ham}), we get a
single-particle gap that varies from one level to another: $\Delta
_{1}^{j}=\bar\lambda M_{jj}/(2\nu _{0})$. The average density of single
electron states in a large sample is controlled by the IPR distribution $%
\mathcal{P}(M)$
\begin{equation}
\nu (\varepsilon )=\nu _{0}\int_{0}^{2\varepsilon \nu_{0}/\bar\lambda}
\mathcal{P}(M)dM  \label{DoS}
\end{equation}
Scaling theory of localization predicts that near the mobility edge $\mathcal{P}
(M)$ acquires a scale-invariant form and this is indeed what was observed in
numerical studies \cite{Mirlin2002}. Moreover, these data indicate that the
distribution $\mathcal{P}(M)$ decreases fast at $M/\bar{M}$ $\rightarrow 0$
so that extended states occur too rare to smear up a gap-like behavior in
the density of states (\ref{DoS}) with the gap value given by Eq.~(\ref%
{DeltaP3}).

We emphasize that the DoS (\ref{DoS}) does not contain any "coherence peak"
above the gap. The same qualitative behavior of the DoS in the insulating region was
obtained \cite{Ghosal2001} by  solving numerically  BCS mean field equations
for the disordered "negative-U" Hubbard model. Evidently, the spectral gap $%
\Delta _{1}$ should be associated with the experimentally determined\cite%
{Shahar1992,Kowal1994} activation energy $T_{I}$. The gap dependence (\ref%
{DeltaP3}) predicts a moderate increase ($\propto 1/L_{\mathrm{loc}}^{D_{2}}$%
) of $T_{I}$ with disorder strength in a complete agreement with
the experimental data \cite{Shahar1992}, see Fig. 2. Note that attempts to fit
the data with a gap that scales as $\delta _{L} \propto 1/L_{\mathrm{loc}%
}^{3}$ have failed spectacularly \cite{Shahar1992}.

\begin{figure}[ht]
\includegraphics[width=2in]{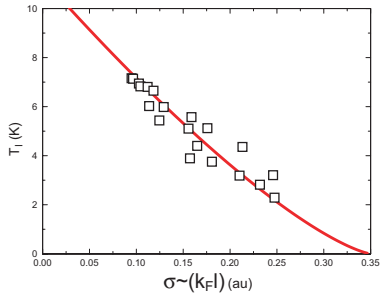}
\caption{Experimental values of the gap, $T_I$ (boxes) and a fit to the
equation (\protect\ref{DeltaP3}) with $L_{loc}\sim (\sigma-\sigma_c)^{-1}$
that involved only the overall scale adjustment \protect\cite{Ovadyahu2005}. }
\label{DataFit}
\end{figure}

We now turn to the parameter region (\ref{cond2}) where one expects a global
superconductive coherence. Namely, the condition $\delta _{L}\ll \Delta$
assures that a large number of eigenstates $\psi _{j}$ with energies down to $%
\epsilon_{j}\sim \Delta$ overlap significantly in the real space, so one can use
the standard BCS-type self-consistent gap equation as first argued in Ref.~%
\onlinecite{Anderson1959}.

However, due to an extra cost of creating
an unpaired electron, the energy required to add an electron in this regime is no longer equal to the self-consistent gap $\Delta $  as in the usual BCS theory. Indeed, because the addition of an electron increases
the number of blocked states in (\ref{grain1}) by one, we have
\begin{equation}
\Delta _{1}=\Delta +\frac{G}{2}\approx\Delta+\frac{\bar\lambda }{%
2\nu_{0}L_{\mathrm{0}}^{3}(L_{\mathrm{loc}}/L_{\mathrm{0}})^{D_{2}}}
\label{Delta1}
\end{equation}

In order to evaluate the energy of pair excitations, i.e. the excitations of
the condensate with no unpaired electrons, it is convenient to rewrite
Hamiltonian (\ref{grain1}) in terms of Anderson's pseudospin-1/2 operators
\cite{Anderson1958} defined on the subspace of unblocked states as $%
s_{j}^{+}=c_{j\uparrow }^{\dagger }c_{j\downarrow }^{\dagger
}=(s_{j}^{-})^{\dagger }$ and $2s_{j}^{z}=c_{j\uparrow }^{\dagger
}c_{j\uparrow }+c_{j\downarrow }^{\dagger }c_{j\downarrow }-1$. We have
\begin{equation}
H=\sum_{j}2\epsilon_{j}s_{j}^{z}-g\sum_{j,k}s_{j}^{+}s_{k}^{-}+\sum_{B}%
\left( \epsilon _{k}+\frac{G}{2}\right)  \label{hspin}
\end{equation}
In the BCS mean-field approximation eigenstates of (\ref{hspin}) correspond
to all spins being parallel or antiparallel to the effective magnetic field $%
B_{j}=-(2\Delta ,0,2\epsilon _{j})$, where $\Delta =g\sum_{j}\langle
s_{j}^{x}\rangle $, leading to a self-consistency condition
\begin{equation}
\frac{2}{g}=\sum_{j}\frac{e_{j}}{\sqrt{\epsilon_{j}^{2}+\Delta ^{2}}}
\label{gap}
\end{equation}
where $e_{j}=1$ if the spin is against the magnetic field and $e_{j}=-1$
otherwise.

In the ground state all spins are along their magnetic fields, $e_{j}=1$ for
all $j$. The lowest pair excitation corresponds\cite{Anderson1958,BCS1957}
to a single spin flip at the Fermi level $\epsilon _{0}=0$ and has an energy
$\Delta _{2}=2\Delta $. It also modifies the gap equation since now one of $%
e_{j}$ becomes negative, $e_{0}=-1$. Because the number of states in the
localization volume $L_{\mathrm{loc}}^{3}$ is finite, this modification
results~\cite{Matveev1997,Yuzbashyan2005} in a negative correction to the
gap: $\delta\Delta = -\delta _{L}/2$. Thus
\begin{equation}
\Delta _{2}=2\Delta _{0}-\delta _{L}=2\Delta _{0}-\frac{c}{\nu _{0}L_{%
\mathrm{loc}}^{3}}  \label{Delta2}
\end{equation}%
here $\Delta_0$ is the $T=0$ gap at large $L_{loc}$, numerical coefficient $%
c=1$ for model (\ref{grain}) and $c\sim 1$ for model (\ref{Ham}). Note that
pair excitations with gap $\Delta_2$ are spinless, whereas single-particle
excitations (gap $\Delta_1$) carry spin $1/2$. Thus $\Delta_1$ is the
\textit{spin gap}, while $\Delta_2$ is the energy gap.

In the regime (\ref{cond2}), the typical single-particle gap $\Delta _{1}$
remains much larger than the gap for pair excitations $\Delta _{2}$. This
leads to the exclusion of single-particle excitations from the
thermodynamics at $T\ll \Delta _{1}$, which results in a modified mean-field result
$T_{c}/ \Delta(T=0)=2[T_{c}/\Delta(T=0)]_{BCS}$. The actual dependence of $T_c$
on the localization length is not easy to find due to enhanced role of thermal
fluctuations near the mobility edge~\cite{Sad1984,Kot85,Ma1985}.

We emphasize that $\Delta (T=0)$ does not characterize the spectral gap in
the present situation. The most naturally measured (e.g. by tunneling
conductance or optical conductivity experiments) spectral parameter is the
single-particle gap $\Delta _{1}$, which according to Eqs.~(\ref{Delta1},\ref%
{cond2}) is much larger than $T_{c}$. An anomalously large ratio $\Delta
_{1}/T_{c}$ leads to the insulating trend of the resistivity versus
temperature behavior in the intermediate temperature range $T_{c}<T\leq
\Delta _{1}$. This was observed in many strongly disordered superconductors
and is especially well-known for underdoped cuprates~\cite{Steiner2004}.

The quantitative similarity between $R(T,B)$ behavior in InO$_{x}$ films and
underdoped cuprates noticed in Ref~\onlinecite{Steiner2004}~(see also Ref.~%
\onlinecite {Sun2005}) makes us believe the pseudogap in underdoped cuprates
may have a similar origin: pairing of electrons on localized states. The
important difference of the cuprates is the d-wave symmetry of the pairing.
We plan to address this issue in future work. Suppression of single-particle
density of states in the pseudogap regime can be observed by measuring the
differential tunneling conductance which we expect to follow the universal IPR
distribution $\mathcal{P}(M)$ studied in Ref.~\onlinecite{Mirlin2002}:
\begin{equation}
\frac{dI}{dV}\propto \left( \frac{d\nu }{d\varepsilon }\right) _{\varepsilon
=eV}\propto \mathcal{P}\left( 2eV\nu _{0}/\bar\lambda  \right)   \label{diff}
\end{equation}%
As emphasized above, the generic features indicating a formation of the
novel insulating state near superconductor-insulator transition (such as the
resistivity maximum as a function of magnetic field or disorder) were
observed in many different materials. However, the details may vary: in InO$%
_{x}$ thick (essentially 3D) films the resistivity above the transition has
an activated behavior while in very thin (2D) films \cite%
{Goldman1993,Baturina2003} it follows an Efros-Shkovskii law. The theory
developed in this Letter is applicable only to 3D systems but it seems
likely that a similar physics leads to a weaker but observable effects in 2D
films.

In conclusion, weak Anderson insulators with Cooper attraction are shown to
possess hard insulating gap whose magnitude is determined by the IPR
statistics near the mobility edge. Although this gap is of a superconducting
origin, it does not lead to a coherence peak. In the ground-state of this
insulator all electrons are paired on individual localized eigenfunctions.
When the Fermi-level approaches the mobility edge, superconductive
correlations develop between localized pairs. Key features of the predicted
superconductive ground-state are an unusually large (compared to $T_{c}$)
single-particle excitation gap (spin gap) and a pseudogaped regime at
temperatures about $T_{c}$. These unusual features stem from the fractal
nature of localized eigenstates near the mobility edge. We assumed that $%
\psi _{j}^{2}(\mathbf{r})$ of the relevant nearly-critical eigenfunctions
have weak mutual correlations. If the correlations are in fact strong \cite%
{Kravtsov}, the description of the insulating phase is not affected but the
ratio $\Delta _{1}/\Delta \gg 1$ will be significantly reduced. Finally, our
theory is based on a phenomenological assumption of an \textit{attractive}
sign of the diagonal coupling constant $\bar{\lambda}>0$. Another
interesting situation is realized when $\lambda >0$, but $\bar{\lambda}<0$.
In this case we expect a formation of "effective magnetic impurities" and
strong suppression of $T_{c}$. We plan to consider this situation separately.

We are grateful to B. L. Altshuler, T. I. Baturina, V. F. Gantmakher, A. S.
Ioselevich, V. E. Kravtsov, A. Millis, A. D. Mirlin, M. Mueller, Z.
Ovadyahu, D. Shahar for useful discussions. This research was supported by
NSF grant DMR 0210575, by NATO CLG grant 979979, by RFBR grants 04-02-16348,
04-02-08159 and by the program "Quantum Macrophysics" of RAS.


\begin{thebibliography}{99}
\bibitem{Shahar1992} D.Shahar and Z.Ovadyahu, Phys. Rev. B \textbf{46},
10917 (1992)

\bibitem{Goldman1993} Y.Liu \textit{et al},
hys. Rev. B \textbf{47}, 5931 (1993).

\bibitem{Kowal1994} D.Kowal and Z.Ovadyahu, Sol.St.Com. \textbf{90}, 783
(1994).

\bibitem{Gantmakher2000} V. F. Gantmakher \emph{et al} Pis'ma ZhETF \textbf{%
71}, 693 (2000).

\bibitem{Adams2001} V. Butko and P. Adams, \emph{Nature}, \textbf{409}, 161
(2001).

\bibitem{Baturina2004} T. I. Baturina \emph{et al}, Pis'ma ZhETF \textbf{79}
416 (2004).

\bibitem{Shahar2004} G. Sambandamurthy \emph{et al}, Phys. Rev. Lett.
\textbf{92}, 107005 (2004).

\bibitem{Shahar2005} G. Sambandamurthy \emph{et al}, cond-mat/0403480.

\bibitem{Steiner2005} M. A. Steiner and A. Kapitulnik, cond-mat/0406227.

\bibitem{Gantmakher1998} V. F. Gantmakher \emph{et al}, JETP Letters \textbf{%
68}, 345 (1998); V. F. Gantmakher, Physics-Uspekhi \textbf{41}, 214 (1998).

\bibitem{Anderson1959} P.W. Anderson, J.Phys.Chem.Solids \textbf{11}, 26
(1959).

\bibitem{Abrikosov1959} A.A. Abrikosov and L.P. Gorkov, Sov. Phys. JETP 8
(1958) 1090.

\bibitem{Sad1984} L. N. Bulaevsky and M. V. Sadovsky, JETP Letters \textbf{39%
}, 640 (1984).

\bibitem{Kot85} A. Kapitulnik and G. Kotliar, Phys. Rev. Lett. \textbf{54},
473 (1985).

\bibitem{Ma1985} S.-K. Ma and P.A. Lee, Phys. Rev. B \textbf{32}, 5658
(1985).

\bibitem{Larkin1999} A.I.Larkin, Ann.Phys. (Leipzig) {\bf 8}, 507 (1999).

\bibitem{Parshin1999} D.A.Parshin and H.R.Schober, Phys. Rev. Lett. \textbf{%
83}, 4590 (1999).

\bibitem{Mirlin2002} A.Mildenberger, F.Evers and A.D.Mirlin, Phys. Rev. B
\textbf{66}, 033109 (2002).

\bibitem{cuprates} K. M. Lang, \emph{et al} Nature, 415 412 (2002).

\bibitem{Matveev1997} K.A.Matveev and A.I.Larkin, Phys. Rev. Lett. \textbf{78%
}, 3749 (1997).

\bibitem{Anderson1958} P. W. Anderson: Phys. Rev. \textbf{112}, 1900 (1958).

\bibitem{BCS1957} J. Bardeen, L.N. Cooper, and J.R. Schriefer, Phys. Rev.
\textbf{108} 1175 (1957).

\bibitem{Yuzbashyan2005} E. A. Yuzbashyan, A. A. Baytin, and B. L.
Altshuler. Phys. Rev. B 71, 094505 (2005).

\bibitem{Ghosal2001} A.Ghosal, M.Randeria and N.Trivedi, Phys. Rev. B
\textbf{65}, 014501 (2001).

\bibitem{Ovadyahu2005} The fit was communicated to us by Z. Ovadyahu.

\bibitem{Steiner2004} M. A. Steiner, G. Bobinger and A. Kapitulnik,
cond-mat/0406232.

\bibitem{Sun2005} X. F. Sun, K. Segawa and Y. Ando, cond-mat/0502223.

\bibitem{Baturina2003} T.I.Baturina \textit{et al}, cond-mat/0309281,
version 1.

\bibitem{Kravtsov} V.E.Kravtsov, private communication.
\end{thebibliography}
\end{document}